\documentclass[conference]{IEEEtran}
\IEEEoverridecommandlockouts
\usepackage[utf8]{inputenc}
\usepackage{cite}
\usepackage{amsmath,amssymb,amsfonts}
\usepackage{algorithmic}
\usepackage{graphicx}
\usepackage{textcomp}
\usepackage{siunitx}
\usepackage{float}

\usepackage[hidelinks]{hyperref}
\usepackage{doi}
\usepackage[utf8]{inputenc}
\usepackage[table,xcdraw,x11names]{xcolor}
\usepackage{siunitx}
\usepackage{soul}
\usepackage{blindtext}
\usepackage{lipsum}
\makeatletter
\@ifpackagelater{lipsum}{2019/01/03}{\setlipsum{auto-lang=false}}{}
\makeatother

\usepackage{ragged2e}
\usepackage{xparse}
\usepackage{multirow}
\usepackage{subcaption}

\usepackage{flushend}

\def\BibTeX{{\rm B\kern-.05em{\sc i\kern-.025em b}\kern-.08em
    T\kern-.1667em\lower.7ex\hbox{E}\kern-.125emX}}

\begin{document}

\title{A Tool for the Synthesis of Adaptive Probabilistic Processors Based on the Ising Model}


\ifx\blindreview\undefined

\author{
\IEEEauthorblockN{
Jonathan Juracy Carneiro da Silva\IEEEauthorrefmark{1}, 
Leonardo R. Gobatto\IEEEauthorrefmark{1}\IEEEauthorrefmark{2}, 
and Jose Rodrigo Azambuja\IEEEauthorrefmark{1}\IEEEauthorrefmark{2}%
}

\IEEEauthorblockA{\IEEEauthorrefmark{1}\textit{Federal University of Rio Grande do Sul (UFRGS) -- Institute of Informatics  -- PGMicro}, Porto Alegre, Brazil\\\{jonathan.silva, leonardo.gobatto, jose.azambuja\}@inf.ufrgs.br\\
}
\IEEEauthorblockA{\IEEEauthorrefmark{2}\textit{Center for Embedded Devices and Research in Digital Agriculture (CEDRA)}, São Leopoldo, Brazil}

}

\else

\author{
   \IEEEauthorblockN{Omitted for blind review.}
   \vspace{-5mm}
   
}

\fi

\maketitle
\thispagestyle{plain}
\pagestyle{plain}

\IEEEpubidadjcol 

\begin{abstract}
This work presents a tool for the synthesis and simulation of probabilistic architectures for solving combinatorial optimization problems by mapping them to the Ising model. The proposed approach automatically constructs the Ising Hamiltonian and determines the number of probabilistic elements (p-bits) based on problem characteristics such as size and topology. Furthermore, the tool introduces an adaptive strategy for selecting the most suitable update algorithm among Gibbs Sampling, Simulated Annealing (SA), Simulated Quantum Annealing (SQA), and cluster-based methods. Experimental results using benchmark problems demonstrate improved convergence behavior and flexibility compared to fixed approaches. The proposed framework enables systematic evaluation of probabilistic computing strategies and supports the development of future hardware implementations based on MTJs and p-bits.
\end{abstract}

\section{Introduction}

Combinatorial optimization problems are ubiquitous in science and engineering, often involving NP-hard formulations that challenge deterministic solvers as instance sizes scale \cite{barahona1982np,lucas2014ising}. In practice, this limits the use of exhaustive methods, especially when the solution space contains many local minima and highly multimodal energy landscapes. Within this context, physics-inspired methods—specifically the Ising model—have emerged as a robust framework \cite{lucas2014ising}. By mapping binary decision variables onto interacting spins, optimization becomes a search for the Hamiltonian's ground state.

While recent advances in probabilistic computing and hardware-oriented units like p-bits have enabled efficient Ising Machines \cite{camsari2017pbits,kaiser2021pbits}, a significant design gap remains. Most existing architectures, including those based on Magnetic Tunnel Junctions (MTJs), assume fixed resources \cite{singh2024cmosmtj,yang2025mtj}. These constraints require the designer to predefine both the number of elements and the update algorithm, even though the ideal configuration is highly dependent on problem-specific features such as connectivity and coupling density. A static setup that performs well for one class of problems may be inefficient or unnecessarily costly for another.

This work addresses this limitation by proposing a tool for the synthesis of \textbf{adaptive probabilistic processors}. The proposed framework automatically maps optimization instances, estimates required hardware resources, and selects the most suitable update strategy among various stochastic and replica-based dynamics \cite{kirkpatrick1983sa,kadowaki1998qa,das2008qa,raimondo2025highperformance}. In doing so, the tool provides a resource-aware environment for probabilistic optimization.

The main contributions of this work are: (i) a unified preprocessing and synthesis flow for Ising-based probabilistic simulation; (ii) an adaptive allocation strategy for probabilistic elements based on structural characteristics of the input instance; and (iii) an automatic algorithm-selection mechanism that chooses among distinct update dynamics according to the expected characteristics of the energy landscape. Experimental results on representative benchmark problems indicate that the adaptive approach improves flexibility and provides consistent solution quality across distinct instance classes.


\section{Related Work}

The literature extensively uses the Ising model as a unifying framework for combinatorial optimization, mapping NP-hard problems such as Max-Cut, SAT, and graph coloring to spin systems in which low-energy states correspond to optimal solutions \cite{lucas2014ising,barahona1982np}. This paradigm shifted the focus toward specialized optimization through physics-inspired dynamics. Concurrently, probabilistic computing introduced p-bits as stochastic binary units \cite{camsari2017pbits,kaiser2021pbits}, enabling Ising Machines that leverage randomness as a computational resource. Recent hardware-oriented designs using MTJs and nanomagnets further demonstrate the scalability of these probabilistic architectures \cite{singh2024cmosmtj,yang2025mtj}.

Various update strategies drive these machines: Gibbs Sampling serves as a local stochastic baseline \cite{geman1984gibbs}. At the same time, Simulated Annealing (SA) remains a robust heuristic for minimization \cite{kirkpatrick1983sa}. More advanced methods, such as Simulated Quantum Annealing (SQA), employ replica-based dynamics via Suzuki--Trotter mapping to handle frustrated instances \cite{kadowaki1998qa,suzuki1976trotter,das2008qa,raimondo2025highperformance}, while cluster-based updates accelerate convergence in correlated systems \cite{swendsen1987cluster,wolff1989cluster}. Despite this diversity, existing works typically rely on fixed resources and predefined dynamics. In contrast, this study proposes an adaptive synthesis approach that scales p-bits and selects update strategies based on the structural characteristics of each problem.

Recent advances in probabilistic Ising machines have demonstrated that architectures employing multiple replicas can significantly improve convergence and robustness in complex optimization problems. An example of such an architecture is presented in Fig. 1, adapted from [10], which combines probabilistic elements with SQA-inspired dynamics. These developments motivate the adaptive synthesis strategy proposed in this work.

\begin{figure}[t]
	\centering
	\includegraphics[width=\columnwidth]{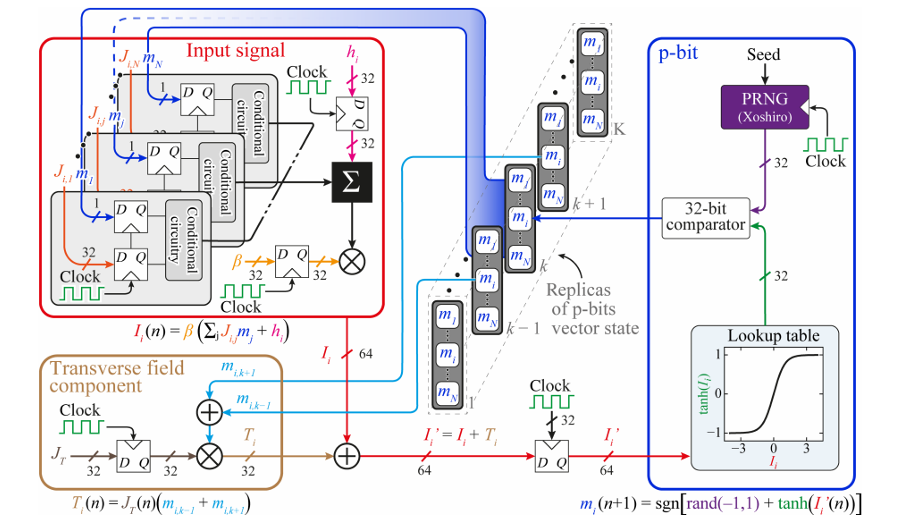}
	\caption{SQA-oriented probabilistic Ising machine architecture, adapted from \cite{raimondo2025highperformance}.}
	\label{fig:raimondo2025_arch}
\end{figure}

\section{Proposed Framework}

The proposed framework was designed as an adaptive probabilistic synthesis flow for combinatorial optimization problems represented as Ising models. Its objective is to provide a unified environment in which the input instance is automatically processed, converted to an Ising formulation, associated with an appropriate number of probabilistic elements, and solved using an update strategy selected based on the problem's structural properties.

\begin{figure}[t]
    \centering
    \includegraphics[width=1\columnwidth, trim=4mm 140mm 4mm 8mm, clip]{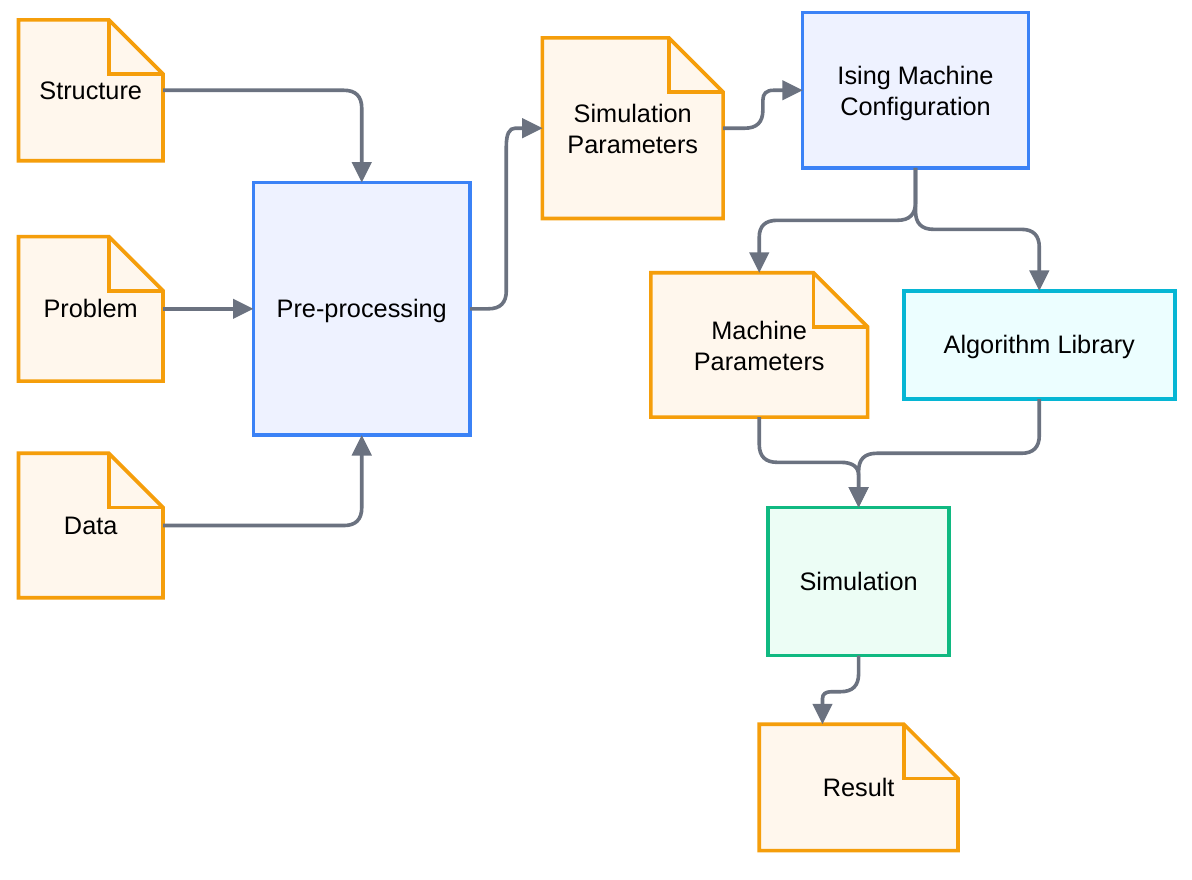}
    \caption{General workflow of the proposed framework.}
    \label{fig:framework_flow}
\end{figure}

The complete workflow is illustrated in Fig.~\ref{fig:framework_flow}. The process starts from the input data, which may represent a weighted graph, a constraint structure, or another combinatorial instance. After validation and standardization, the framework extracts the problem structure and converts it into an Ising representation defined by a coupling matrix $J$ and a local bias vector $h$. In this stage, the optimization objective is reformulated as the minimization of the Ising energy function:

\begin{equation}
E(s) = - \sum_{i<j} J_{ij}s_is_j - \sum_i h_i s_i ,
\label{eq:ising_energy}
\end{equation}

where $s_i \in \{-1,+1\}$ denotes the spin state of the $i$-th probabilistic element.

After the Ising model is constructed, the framework performs a preprocessing step to extract structural descriptors from the instance. These descriptors include the problem size, effective connectivity, and coupling density. Based on this information, the framework estimates the number of probabilistic elements required for simulation. This adaptive allocation is intended to approximate the behavior expected in future hardware-oriented probabilistic architectures, where the available number of p-bits or stochastic magnetic devices must be matched to the complexity of the target problem.

Once the probabilistic resources are defined, the framework configures the Ising machine parameters. It selects an update strategy from Gibbs Sampling, Simulated Annealing (SA), Simulated Quantum Annealing (SQA), or cluster-based updates. This selection is not fixed in advance; instead, it depends on the structural profile of the mapped instance and the expected exploration requirements of the corresponding energy landscape. In this sense, the framework combines algorithmic flexibility with hardware-inspired resource awareness.

The final stage consists of the probabilistic simulation itself, followed by structured result generation. During execution, the framework monitors quantities such as best energy, convergence iteration, execution time, and synthetic cost-related metrics. This organization enables the proposed tool to operate not only as an optimization simulator but also as a synthesis-oriented platform for studying adaptive probabilistic processors within a unified Ising-based formulation.

\section{Experimental Setup and Execution Flow}

The proposed framework was evaluated through a set of simulation campaigns designed to investigate how adaptive probabilistic synthesis behaves under distinct classes of combinatorial optimization problems. In contrast to hardware-oriented experimental studies, where performance is assessed under physical operating conditions, the present work focuses on a software-level evaluation of Ising-based probabilistic processing, aiming to reproduce the decision flow expected in future hardware-aware architectures \cite{lucas2014ising,barahona1982np,camsari2017pbit_prx,kaiser2021pbits}.

The benchmark set was selected to cover a range of graph structures, coupling densities, and energy landscapes. In particular, the experiments included Traveling Salesman Problem (TSP), graph coloring, SAT, matching, segmentation, and Max-Cut instances, each mapped to an Ising representation with problem-dependent coupling matrices and local fields. This choice allows the framework to be tested under both sparse and dense interaction patterns, as well as under configurations with distinct levels of frustration and multimodality, which are known to strongly affect the behavior of stochastic optimization dynamics \cite{lucas2014ising,barahona1982np}.

For each benchmark, the framework automatically determined the number of probabilistic elements based on the instance's structural characteristics, particularly its size and effective coupling density. This adaptive allocation was intended to approximate the behavior of resource-constrained probabilistic hardware, in which the number of available p-bits or stochastic magnetic elements must be matched to the complexity of the target problem \cite{camsari2017pbit_prx,singh2024cmosmtj,yang2025mtj}. In parallel, the tool selected one update strategy among Gibbs Sampling, Simulated Annealing (SA), Simulated Quantum Annealing (SQA), and cluster-based updates, according to the expected difficulty of the energy landscape and the connectivity profile of the mapped instance \cite{kirkpatrick1983sa,geman1984stochastic,kadowaki1998qa,das2008qa_review,swendsen1987cluster,wolff1989cluster}.

After the benchmark instances were mapped into Ising form and the corresponding structural descriptors were extracted, the execution engine processed each problem through an adaptive probabilistic simulation flow. Each run combined three coupled decisions: the Ising representation itself, the number of probabilistic elements assigned to the problem, and the update dynamics selected for state evolution. The execution engine operates on top of a common probabilistic state representation, so that all update methods act on the same mapped Hamiltonian. This design isolates the influence of the selected algorithm from preprocessing decisions, enabling fair comparisons across different dynamics under equivalent problem conditions.

The four update strategies considered in this work were Gibbs Sampling, Simulated Annealing (SA), Simulated Quantum Annealing (SQA), and cluster-based updates. Gibbs Sampling was used as the simplest local stochastic baseline, while SA introduced a temperature schedule to improve exploration of the solution space. SQA extended this idea by incorporating a replica-based mechanism inspired by the Suzuki--Trotter mapping, allowing the framework to approximate quantum-like transitions across the energy landscape. Cluster-based updates were included to test whether collective spin transitions could accelerate convergence in more regular or highly correlated instances \cite{geman1984stochastic,kirkpatrick1983sa,kadowaki1998qa,suzuki1976trotter,das2008qa_review,swendsen1987cluster,wolff1989cluster}.

To preserve comparability across problem classes, all simulations were executed within a fixed iteration budget, with identical reporting criteria across all runs. During execution, the framework recorded the best energy reached, the iteration in which convergence was approximately achieved, the total execution time, and synthetic cost-related metrics associated with power and energy consumption. These measurements jointly capture solution quality, computational effort, and the hardware-oriented implications of adaptive probabilistic synthesis. In this sense, the adopted procedure is aligned with recent evaluations of probabilistic Ising machines, in which algorithmic schedules and device-inspired constraints are analyzed together rather than in isolation \cite{grimaldi2022experimental,raimondo2025highperformance,yang2025mtj}.

For each benchmark problem, the simulation engine's outputs were stored in a structured format for subsequent comparison. This allowed the analysis to be based not only on isolated best-case outcomes, but also on the broader relationship between adaptive allocation, selected update dynamics, and optimization behavior. Together, these simulation campaigns provide the experimental basis for evaluating the proposed framework as a synthesis-oriented tool for adaptive probabilistic processors based on the Ising model.

\section{Results and Discussion}

The proposed framework was evaluated using representative benchmark problems with distinct structural properties, including TSP, graph coloring, SAT, matching, segmentation, and Max-Cut. Table~\ref{tab:energy_full} reports the energy values obtained by each update strategy for all benchmark problems. Unlike a single best-case summary, this representation preserves the comparative behavior of Gibbs Sampling, Simulated Annealing (SA), Simulated Quantum Annealing (SQA), and cluster-based updates across different instance classes.

\begin{figure}[t]
    \centering
    \includegraphics[width=1\columnwidth]{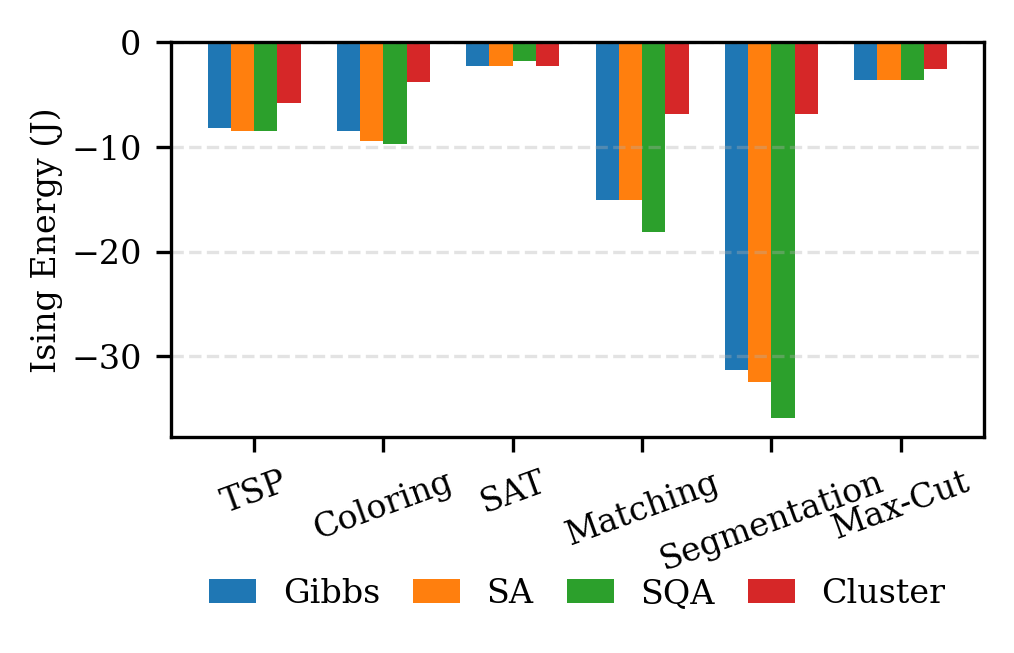}
    \caption{Ising energy comparison.}
    \label{fig:energy_comparison}
\end{figure}

\begin{table}[t]
\caption{Best energy (J) obtained by each update strategy.}
\label{tab:energy_full}
\renewcommand{\arraystretch}{1.3}
\centering
\begin{tabular}{lrrrr}
\hline
Problem & Gibbs & SA & SQA & Cluster \\
\hline
TSP          & -8.19  & -8.50  & -8.49  & -5.81 \\
Coloring     & -8.44  & -9.38  & -9.67  & -3.75 \\
SAT          & -2.30  & -2.30  & -1.81  & -2.30 \\
Matching     & -15.10 & -15.10 & -18.15 & -6.87 \\
Segmentation & -31.29 & -32.44 & -35.91 & -6.89 \\
Max-Cut      & -3.57  & -3.57  & -3.57  & -2.59 \\
\hline
\end{tabular}
\end{table}

As shown in Table~\ref{tab:energy_full}, the relative behavior of the update strategies depends strongly on the benchmark structure. In particular, SQA achieved the lowest energy values in graph coloring, matching, and segmentation, while SA provided the best result in the TSP instance. For SAT and Max-Cut, the difference was smaller, suggesting that simpler dynamics may still suffice in less demanding instances.

Note that the energy values reported in Fig.~\ref{fig:energy_comparison} are negative. This behavior is expected and follows directly from the Ising Hamiltonian formulation used in this work. Due to the negative sign in the Hamiltonian, configurations that satisfy the problem constraints tend to minimize the energy, often resulting in negative values.
In this context, lower energy values indicate better solutions, as they correspond to configurations closer to the system's ground state. Therefore, the goal of the optimization process is not to achieve positive values, but rather to minimize the energy, regardless of its sign.

This explains why methods such as SQA produce more negative values in structured problems like matching and segmentation, reflecting their ability to explore complex energy landscapes better and escape local minima.
In addition to the Ising energy analysis, it is important to evaluate the computational cost of each method. For this purpose, the computational energy is estimated as

{\setlength{\abovedisplayskip}{2pt}
\setlength{\belowdisplayskip}{2pt}
\setlength{\abovedisplayshortskip}{2pt}
\setlength{\belowdisplayshortskip}{2pt}
\begin{equation}
E_{\mathrm{comp}} = Pt
\end{equation}
}

where $P$ is the average power consumption of the processing unit and $t$ is the execution time.

In this work, a constant average power of 65~W was assumed, corresponding to a typical CPU operating condition. Although this is an approximation, it provides a consistent basis for comparing the relative energy efficiency of the evaluated algorithms.

The results show that, although methods such as SQA tend to achieve lower Ising energy values (better solutions), they also incur higher physical energy consumption due to increased execution time. This highlights an important trade-off between solution quality and computational cost.

\begin{figure}[t]
    \centering
    \includegraphics[width=1\columnwidth]{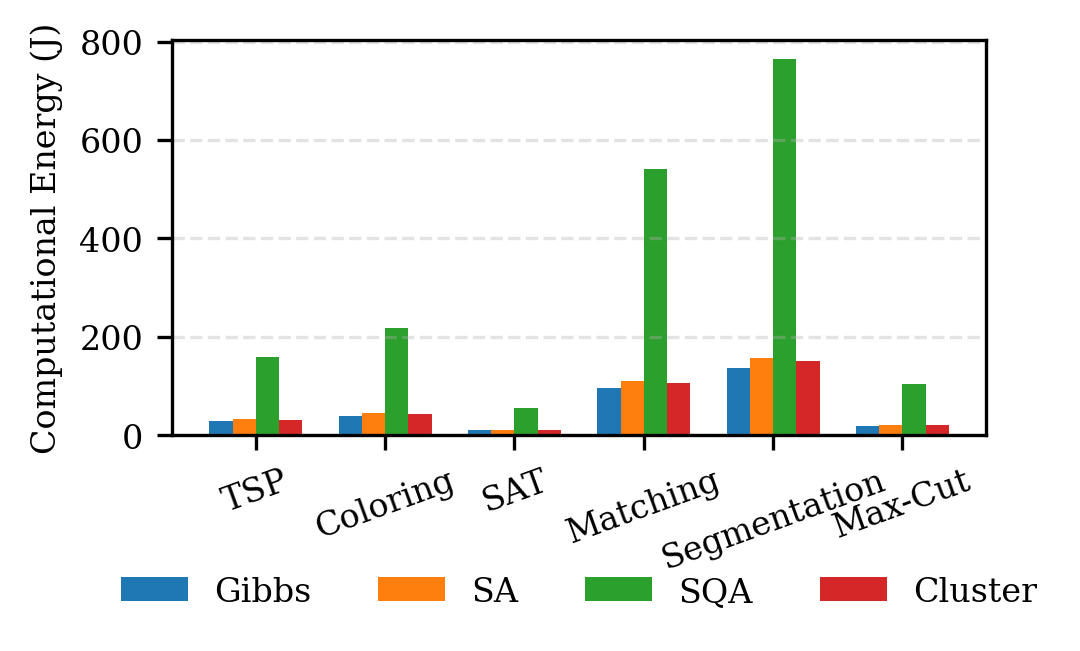}
    \caption{Computational Energy comparison.}
    \label{fig:energy_to_solution}
\end{figure}

Fig.~\ref{fig:energy_comparison} highlights this tendency more clearly. Simulated Quantum Annealing (SQA) consistently reached lower energy values in the most structured and constrained problems, suggesting that replica-based dynamics are advantageous in energy landscapes with stronger frustration and deeper local minima. In contrast, SA remained highly competitive across multiple instances, confirming its role as a robust compromise between exploration capability and practical efficiency.

\begin{figure}[!t]
    \centering
    \includegraphics[width=1\columnwidth]{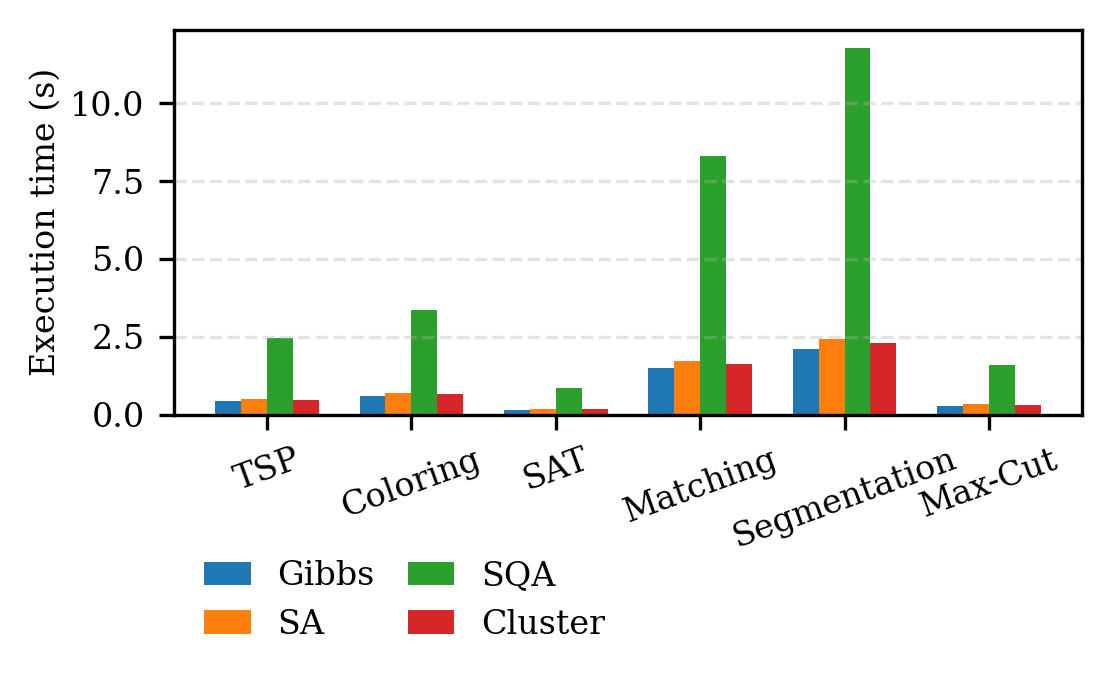}
    \caption{Execution time comparison.}
    \label{fig:time_comparison}
\end{figure}

The execution-time behavior shown in Fig.~\ref{fig:time_comparison} reveals the main trade-off of the evaluated strategies. Although SQA often achieved the best energies, it also required significantly longer execution times, especially in matching and segmentation. This confirms that improved solution quality is obtained at the cost of higher computational overhead. By comparison, SA provided a more balanced profile, while Gibbs Sampling remained efficient in smaller or less complex instances.

\begin{figure}[t]
    \centering
    \includegraphics[width=1\columnwidth]{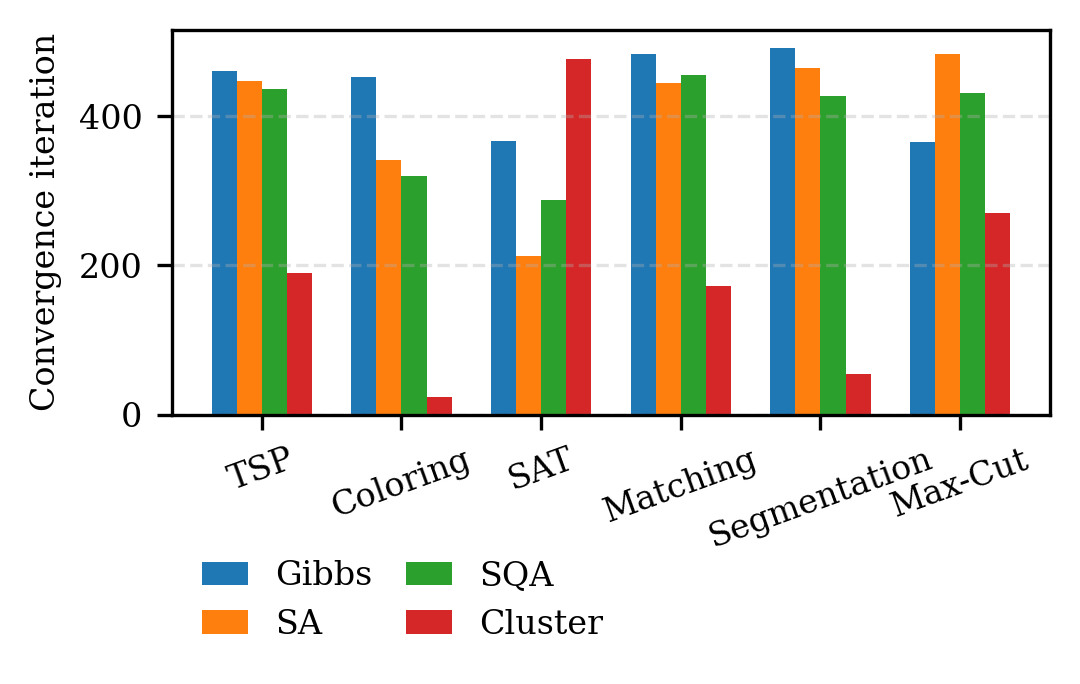}
    \caption{Convergence iteration comparison.}
    \label{fig:convergence_comparison}
\end{figure}

Fig.~\ref{fig:convergence_comparison} shows that fast convergence does not necessarily imply better final energy. Cluster-based updates frequently converged earlier than the other strategies. Still, their final energies were usually less competitive for problems with heterogeneous couplings and stronger constraint penalties. This suggests that collective spin updates are more useful as acceleration mechanisms in regular topologies than as universal optimization strategies for all Ising-mapped problems.

Overall, the results support the main premise of this work: different problem classes benefit from different update dynamics, and therefore adaptive synthesis provides a more coherent approach than fixed probabilistic configurations. In addition, the resource-allocation mechanism remained compatible with each benchmark's complexity, indicating that the proposed framework can serve as a useful basis for future hardware-oriented implementations of adaptive probabilistic processors.

\section{Conclusion}

This work introduced a synthesis tool for adaptive probabilistic processors that adjusts the execution flow to the structural characteristics of each instance. By integrating automatic mapping, adaptive resource allocation, and dynamic algorithm selection, the framework demonstrated coherent decision-making across benchmarks like TSP, SAT, and Max-Cut. Experimental results confirmed that different update strategies entail distinct trade-offs between solution quality and computational cost, validating the need for adaptive selection. 
Furthermore, the resource-allocation strategy aligns the framework with the physical constraints of future hardware architectures based on p-bits and MTJs. Ultimately, the proposed tool bridges algorithmic evaluation and hardware-inspired design. Future developments will focus on expanding benchmark sets and integrating the flow with experimental hardware platforms.


\section{Acknowledgements}
This work was partially funded by Coordenação de Aperfeiçoamento de Pessoal de Nível Superior - Brasil (CAPES) - Finance Code 001, CNPq, FAPERGS Techfuturo - 23/2551-0002199-4, and by the Center for Embedded Devices and Research in Digital Agriculture (CEDRA) of SENAI-RS, with financial resources from the PPI IoT/Manufatura 4.0 / PPI HardwareBR of the MCTI, grant number 056/2023, signed with EMBRAPII.

	\bibliographystyle{IEEEtran}
	\bibliography{biblio}
\end{document}